\documentclass[conference]{IEEEtran}
\usepackage{graphicx}
\usepackage{listings}
\usepackage{xcolor}
\usepackage{amssymb}
\usepackage{amsmath}
\ifCLASSINFOpdf
\else
\fi

\begin{document}

%
\title{Towards Deep Federated Defenses Against Malware in Cloud Ecosystems} 

\makeatletter
\makeatother
\author{\IEEEauthorblockN{
Joshua Payne\IEEEauthorrefmark{1},
Ashish Kundu\IEEEauthorrefmark{2}
}

\IEEEauthorblockA{\IEEEauthorrefmark{1}Stanford University, CA, USA Email: jfp@cs.stanford.edu}
\IEEEauthorblockA{\IEEEauthorrefmark{2}IBM Thomas J. Watson Research Center, Yorktown Heights, NY, USA Email: akundu@acm.org}
}


\maketitle

\begin{abstract}
In cloud computing environments with many virtual machines, containers, and other systems, an epidemic of malware can be highly threatening to business processes. In this vision paper, we introduce a hierarchical approach to performing malware detection and analysis using several recent advances in machine learning on graphs, hypergraphs, and natural language. We analyze individual systems and their logs, inspecting and understanding their behavior with attentional sequence models. Given a feature representation of each system's logs using this procedure, we construct an attributed network of the cloud with systems and other components as vertices and propose an analysis of malware with inductive graph and hypergraph learning models. With this foundation, we consider the multicloud case, in which multiple clouds with differing privacy requirements cooperate against the spread of malware, proposing the use of federated learning to perform inference and training while preserving privacy. Finally, we discuss several open problems that remain in defending cloud computing environments against malware related to designing robust ecosystems, identifying cloud-specific optimization problems for response strategy, action spaces for malware containment and eradication, and developing priors and transfer learning tasks for machine learning models in this area.
\end{abstract}


%
\IEEEpeerreviewmaketitle
\section{Introduction}\label{intro}
Malware is broadly defined as a malicious software program that is intentionally designed to cause damage to a computer by exploiting vulnerabilities in the system. Malware can cause particularly significant disruptions incidents in cloud systems, which contain many virtual machines, containers, and other components, because the instances in these cloud computing environments are often highly interconnected with high-risk trust assumptions and protection mechanisms that are not difficult to break. From a customer's standpoint, this kind of incident can be catastrophic, as malware attacks often lead to the leakage of sensitive data and/or extended downtime of services. The resulting damage is often highly costly and sometimes impossible to fully recover from.

It has been observed that malware often spreads in a behavior similar to that of a biological virus \cite{mishra2010seiqrs}. Once malware infects a host, it is able to use it as a launchpad to other hosts which it seeks to compromise. Hence, in environments with multiple interacting hosts, the impact can be combinatorially large. Identification of malware is currently largely manual, with responses and tactical actions being slow due to the bureaucratic nature of cloud management. Powerful, adaptive, and predictive methods for analyzing the presence and spread of malware in the cloud have yet to be proposed, especially those which utilize knowledge from multiple disjoint and nondisclosing clouds jointly.

In this paper, we propose several methods from the system level up to the multicloud level that are designed to understand and combat malware in an adaptive and holistic manner. Our contributions are as follows: 
\begin{enumerate}
    \item We propose the use of attentional language models for analysis of system logs to featurize their respective systems in a standardized manner for downstream processing.
    \item We view the detection and analysis of malware in cloud as a graph and hypergraph learning problem, proposing several methods for performing inference in a useful way with respect to scores such as risk, exploitability, and impact for individual systems as well as for the cloud as a whole. We also discuss potential approaches for tactical decision-making for managing malware.
    \item We consider the multicloud case, where multiple untrusting clouds may cooperate to learn about the state of malware without divulging private or sensitive information. We propose the use of federated learning to achieve this objective.
    \item We finally pose and discuss several important and difficult open problems related to combatting malware in cloud ecosystems.
\end{enumerate}
As this is a vision paper, we do not dive into great depth with the proposed methods; rather, we offer suggestions for a class of methods that can be used to solve the particular problem. Method and ablation studies are left as future work.

The rest of the paper is organized as follows. In Section \ref{background}, we provide mathematical background on graphs and hypergraphs (the objects we study in the context of the cloud) as well as transductive and inductive learning (important machine learning strategies that we leverage in our approaches). In Section \ref{related}, we explore relevant related work as the foundation from which we build our own. This work is broadly categorized into general malware detection, graph-based and hypergraph-based machine learning, natural language models, and federated learning. In Section \ref{problem}, we outline the problem we aim to solve as well as an overview of how we envision the solution. In Section \ref{proposed}, we dive deeper technically into how these problems are solved, organizing the overall problem statement into three levels: system-level, cloud-level, and multicloud-level analysis. in Section \ref{open}, we pose and discuss several open problems in this area. Finally, in Section \ref{conclusion}, we summarize and recapitulate the contributions of this work.

\section{Background}\label{background}
\subsection{Graphs and Hypergraphs}\label{app:bg_hgraphs}
A \textit{hypergraph} $\mathcal{H}$ = $(V,E)$ is comprised of a finite set of vertices $V = \{v_1, v_2, \ldots, v_n\}$ and a set of hyperedges $E = \{e_1, e_2, \ldots, e_m\} \subseteq 2^V$. We consider connected hypergraphs with $|V| \geq 2$. A \textit{graph} $\mathcal{G}$ is a hypergraph where $|e_i| = 2$ for each $e_i\in E$. Graphs are well-studied in the field of machine learning, but are not capable completely representing the information captured in hypergraphs generally. This fact has practical implications as well: for instance, one cannot represent a multicloud environment with shared systems or hosts using pairwise relationships, as the systems may function with each other differently in different contexts. These hyperedges are said to be \textit{indecomposable} \cite{dhne}. Hence, while the theory built for studying graphs can be utilized to some capacity in the hypergraph context, it is more effective to use hypergraph learning approaches on problems involving hypergraphs.

\subsection{Transductive and Inductive Learning}\label{app:bg_learning}
In \textit{transductive} (or \textit{semi-supervised}) inference tasks, one often seeks to learn from a small amount of labeled training data, where the model has access to labeled and unlabeled data at training time. Formally, we have training instances $\{\textbf{x}_i\}_1^n$ where $\{\textbf{x}_i\}_1^l$ are labeled instances and $\{\textbf{x}_i\}_{l+1}^{l+u}$ unlabeled instances, and corresponding labels $y_i$ in $\{y_i\}_1^n$. Our aim is to learn a function $F : \{\textbf{x}_i\}_1^n \rightarrow \{y_i\}_1^n; \ \textbf{x}_i \mapsto y_i$. Typically, in the case of transductive learning on graphs and hypergraphs, we seek to leverage topological information to represent the vertices in some continuous vector space $\mathbb{R}^{d}$ by embeddings which capture the vertices' or hyperedges' context (homophily and structural equivalence). As such, in the pre-training procedure for finding vertex embeddings, we want to find an embedding function $\Phi : V \rightarrow \mathbb{R}^d$ that maximizes the likelihood of observing a vertex in the sampled neighborhood $N(v)$ of $v$ given $\Phi(v)$: \begin{equation*}
 \max_\Phi \displaystyle\sum_{v\in V}\log \mathbb{P}(N(v) \ | \ \Phi(v))
\end{equation*}
The procedure for finding contextual embeddings for hyperedges is similar. Once we've learned $\Phi$, we can use the embeddings $\Phi(v)$ to learn $F$ in our transductive learning procedure.

In \textit{inductive} (or \textit{supervised}) inference tasks, we are given a training sample $\{\textbf{x}_i\}_1^n \subseteq X$ to be seen by our model, and we want to learn a function $g : X \rightarrow Y; \ \textbf{x}_i \mapsto y_i$ that can generalize to unseen instances. This type of learning is particularly useful for dynamic graphs and hypergraphs, when we may find unseen vertices as time passes, or when we want to apply transfer learning to new graphs and hypergraphs altogether. Here, the representation learning function $\Phi$ is typically dependent on the input features $f(v)$ for a vertex $v$.

\section{Related Work}\label{related}
\subsection{Malware Detection}
Malware spreading behaviors have been likened to viral epidemics in \cite{yu2015malware, mishra2010seiqrs}, lending intuition that informs our own approach. Insight into tracking of API call sequences recursively and understanding behavior to detect and unpack malware samples present in the system was discovered in \cite{ndibanje2019cross}. A non-linear malware propagation paradigm and design of control strategy based on Pontryagin's maximum principle was proposed in \cite{zhu2015dynamical}. Various hardware implementations that perform deep packet inspection to find malicious payloads have also been proposed \cite{hale2012network}.  In \cite{watson2015malware}, an SVM-based method for performing anomaly detection was proposed for detecting malware in cloud. We extend this technique with more powerful models and make use of network topology alongside system logs and other metadata to perform inference.

In 2017, DeepLog, an LSTM-based architecture for performing inference and other natural language tasks specifically on system logs, was proposed in \cite{du2017deeplog} by Du et al. DeepLog captures potentially nonlinear dependencies among log entries from training data that correspond to normal system execution paths, and when a log is generated from a malware-infected system, the hope is that the resulting representation will appear far out of the distribution.

\subsection{Learning from Graphs and Hypergraphs}
 In 2013, Mikolov et al. proposed an unsupervised learning procedure, \textit{skip-gram}, in \cite{word2vec} which uses negative sampling to create context-based embeddings for terms given sentences in a corpus of text. This procedure was used by Perozzi et al. in \textit{DeepWalk} \cite{deepwalk} to treat random walks on a graph as ``sentences" and vertices as ``terms", which outperformed existing spectral clustering and weighted vote-based relational neighbor classifiers \cite{spectral, ane, tang3,latent,rel}. Similar random walk-based approaches followed, such as random walks with bias parameters \cite{node2vec}, methods that utilize network attributes as additional features \cite{planetoid, DANE, tadw} and approaches that could be extended to perform inductive learning \cite{planetoid, graphsage}. Graph convolutions were formally defined by Bruna et al. \cite{bruna} and elaborated upon by Kipf and Welling in \cite{gcn}, who proposed \textit{graph convolutional networks}. Graph convolutions have also been used in variational graph autoencoding in \cite{vgae}. A number of other important graph neural network approaches have very recently been proposed, including GraphRNN \cite{graphRNN}, which gives a deep recurrent generative model for graph generation; Graph Attention Networks \cite{GAT}, which incorporate masked self-attentional layers which allow vertices to attend to their neighbors' features, specifying different weights to different vertices in the neighborhood; and Graph Isomorphism Networks \cite{GIN}, which replaces the MEAN aggregator over nodes, such as the one in GCN, with a SUM aggregator, and add more fully-connected layers after aggregating neigboring node features. This idea is also applicable to hypergraphs. Hypergraph learning is lesser-studied, but a variety of approaches have nonetheless been proposed. In 2007, Zhou et al. proposed methods for hypergraph clustering and embedding in \cite{zhou}, but these methods incur high computational and space complexity. Random walks on hypergraphs have been established, and have likewise been demonstrated as useful in inference tasks \cite{chitra,dwalk,hyperedge2vec}, but these methods do not directly account for the set membership and contextual properties of hyperedges simultaneously and efficiently. Very recently, hypergraph convolution and attention approaches have been proposed \cite{hgnn, sbai} which define a hypergraph Laplacian matrix and perform convolutions on this matrix. Deep Hyperedges was recently proposed as a unified framework for performing inference on hyperedges and vertices using membership and contextual properties jointly \cite{dhe}.

\subsection{Language Models for System Log Analysis}
LSTMs \cite{lstm} (Long Short Term Memory) are a class of recurrent neural network (RNN) which processes and creates representations of sequential data. DeepLog \cite{du2017deeplog} uses and LSTM architecture to analyze system logs for abnormalities and perform other tasks.

In 2018, BERT (Bidirectional Encoder Representations from Transformers) \cite{bert} was demonstrated to outperform previous language models such as LSTMs on a variety of tasks with high efficiency. Other works also explore multi-task learning on language models that leverage large amounts of cross-task data, which also exhibit from a regularization effect that leads to more general representations to help adapt to new tasks and domains \cite{liu2019multi}. Moreover, recent work and tools have shown that interpretability is tractable with BERT \cite{Clark_2019, Hoover2019exBERTAV}.
 
BERT builds upon the encoder structure of the encoder-decoder architecture of transformer, which uses multi-head self attention \cite{bahdanau2014neural, bau2018identifying, transformer}. Different techniques have been developed to better interpret attention maps, such as attention matrix heatmaps \cite{bahdanau2014neural, rush2015neural, rocktaschel2015reasoning} and bipartite graph representations \cite{liu2018visual, strobelt2018s}. This can assist in explainability and interpretability. 

BERT has thus far been used primarily in natural language tasks, and we propose its use in the domain of system logs, specifically. This may simply require a fine-tuning task, or could entail a retraining of BERT from scratch.

\subsection{Federated Learning}
Distributed machine learning is an important concept that has been well-studied \cite{distML} and was an important factor in the development of federated learning \cite{fl,fl2, mcmahan2016communicationefficient}, a system which imposes a more structured approach fitting to a specific domain which has certain constraints such as much lower bandwidth and reliability compared to federated/central nodes. This protocol may not allow for arbitrary distributed computation. Moreover, federated learning can provide an important guarantee of privacy of the training data---participating model trainers may contribute to the training of the model without needing to divulge specific information they have. Federated learning has mostly thus far been explored in the context of mobile devices; we instead propose its utility in the context of cloud computing environments and clusters.
\section{Problem Definition and Vision}\label{problem}
Attackers are always developing new malicious software \cite{statistics2017trends}. Some common examples include: 
\begin{enumerate}
    \item DDoS attacks: a botnet is leveraged to make rapid queries to a service, effectively shutting it down or causing an increase in latency.
    \item Hypercall attacks: an attacker uses a virtual machine to exploit the victim's Virtual Machine Manager (VMM) hypercall handler, perhaps giving the attacker the ability to run arbitrary code.
    \item Hypervisor DoS: an attacker uses a large amount of the hypervisor’s resources to exploit design flaws.
    \item Man in the Middle (MITM): an attacker eavesdrops upon and perhaps modifies messages between two communicators.
    \item Hyperjacking: an attacker attempts to assume control of the VM's hypervisor, giving them access to the entire machine.
    \item Co-Location: an attacker attempts to find the host location of a VM and place their own VM alongside it. Once successful, the attacker can perform cross side-channel attacks.
    \item Live Migration Attack: when VMs are migrated between cloud services, attackers can trick the service into creating multiple migrations, leading to DoS attacks.
\end{enumerate}

Attackers are adaptive, and in order to compete, we must have a system that can adapt in kind to resist malware. These kinds of attacks have varying levels of sophistication and have different characteristics of attack and spread. We've realized that the infection and propagation of malware is inevitable in many systems, and have concluded that in these cases it is always best to contain malware to limited number of hosts without disruption of business processes in the cloud.

Broadly, we want to propose a system that can detect the presence of malware in one or more hosts or systems in a cloud environment and take an action to prevent the spread of malware. When an infected host has been detected in cloud, we want to be able to build a profile of all the connected hosts and devise a containment strategy. To most effectively perform this containment, we need to be able to identify nodes that are more influential in the spread of malware and perform a preventative action that maximizes the probability of completely containing the malware while minimizing the negative effects of downtime, latency, etc. on business processes. 

Currently, to execute actions successfully, multiple stake-holders need to be involved. Typically, these include a cloud system administrator who may decide which port to shutdown of an affected host, a network security administrator who may decide which switch/router to shutdown, and a compliance officer/client security focal who asses risk involved in shutting down various parts of system. This work can be resource-consuming, unreliable, and slow; hence there remains the need for a platform that helps to codify policies in the form of rules which automates resolution of intrusion incident.

To more concretely define our problem, we adopt the scoring strategy proposed in \cite{attackcircuits}. Given a cloud of hosts and systems and any additional knowledge about them (e.g., output logs), we want to be able to compute scores that describe the various security aspects of a cloud in the context of a malware infection. Examples of these would be a risk score, exploitability score, and impact score. Risk is meant to be interpreted as a holistic measure of the security state of the system, network, or cloud, which evaluates the confidentiality, integrity, and availability risks of the object's potential vulnerabilities. Exploitability is a measure of how difficult it would be for an adversary to compromise the object. This can be evaluated at the system level, or by looking at network topology (how central is the vertex in the network?). Impact is a measure of the level of harm or compromise an adversary could inflict in the case of malware infection. Intuitively, nodes with greater centrality or higher degree would have higher risk scores, as well as those with sensitive or important information. 

Given these scores, we would like to perform actions that contain the spread of malware in the most effective way possible. Effectiveness is measured as a function of time-to-implement for the decision, lowest negative impact on business processes, and probability that malware is contained given an implementation of the decision.
\section{Proposed Methods}\label{proposed}
We approach this problem hierarchically, starting from the bottom (analyzing the systems themselves) and working our way to the top (investigating structures that involve multiple clouds, referred to as multicloud environments). At the lowest level, we consider a single system or host. This is an object that has certain properties from which we may draw inference as to its security state and malware status. We will refer to these systems as vertices. In the level above, we consider a group of these devices, linked together in a network. This may be considered as a single cluster, network, or cloud---for our purposes, the algorithms proposed in this section will apply to any of these. We consider the implications of dealing with private, public, and hybrid clouds in this layer. At the top layer, we consider a network of networks or collection of clouds: the multicloud environment. Here, we observe multiple cloud environments interacting with each other in an trusted or untrusted manner, where systems and information may or may not be shared between clouds.

\subsection{Level I: System Analysis}
Within a cloud infrastructure level we consider the elements which are hardware servers that run hypervisors to host virtual machines (VMs). There also exist network infrastructure elements which provide the connectivity within cloud and to external service users. We refer to each of these elements as systems individually, or vertices when considering the rest of the network. The first step to analyzing the presence and spread of malware in the cloud is understanding the systems contained within the cloud which may be infected by or susceptible to malware. 

Each system some level of information associated through it---for example, its hardware infrastructure, operating system (OS), indegree and outdegree neighborhood (other systems it sends and receives messages to and from), and system logs. We can represent some of these properties easily in standardized vectors; however, system logs in particular are information-rich and are not easy to represent uniformly as a sequence of low-dimensional vectors. In 2017, Du et al. \cite{du2017deeplog} proposed an LSTM-based model for anomaly detection to record system states and significant events at various critical points and to help debug system failures and perform root cause analysis. This system, DeepLog, automatically learns log patterns from normal execution detects anomalies when log patterns deviate from the model trained from log data under normal execution. Naturally, this idea can be extended to malware detection: when devices are infected with malware, the system logs should demonstrate abnormalities that can be identified by a natural language sequence model. Around the same time, a novel natural language architecture called the Transformer was proposed \cite{transformer} which outperformed LSTMs and RNNs on many benchmark tasks, making use of an attentional layer. BERT was introduced a year later, adding a pretraining task which involves masking some percentage of the tokens and creating context-capturing representations by defining loss on the prediction of these masked tokens. 

We propose the use of BERT or other attentional model for creating a consummable representation of system logs. Formally, let $v$ be a vertex, or system, in our cloud. This element has associated with it a sparse token sequence representation of logs $\texttt{logs}(v)$, taken from the set of tokens $T$ which we would like to represent compactly and meaningfully as a vector $r\in \mathbb{R}^d$, for $d$ small. An initial, efficient approach to generate a representation of the logs of a system would be to average the representations of its tokens generated by BERT or other sequence model: $$r = \frac{1}{|\texttt{logs}(v)|}\displaystyle\sum_{t \in \texttt{logs}(v)}f_{NL}(t)$$ where $f_{NL}:T \rightarrow \mathbb{R}^{d}$ is a function learned by pretraining via masking on logs, as well as optionally fine-tuning on some downstream task. Details can be found in \cite{bert}. Fine-tuning tasks may include malware-infected system logs determination if the datasets are available; operating system category for better differentiation based on OS, latency regression, et cetera.

There exist more expressive and computationally intensive methods for representing the logs as a standardized vector $r$. Two such approaches are Doc2Vec \cite{le2014distributed} and Doc2VecC \cite{chen2017efficient}. In Doc2Vec, one can use the distributed bag-of-words model, where a single word is to be predicted from its context and context words are the preceding words, or distributed memory, where the task is to predict a single context word using only the document vector. In Doc2VecC, the document is corrupted; that is, some percentage of words are removed, making the task more challenging but also reducing the training runtime.

An initial, bottom-level approach to analyzing malware would be to perform inference simply on the document representation of the logs of a system. This is essentially what DeepLog proposes, but with the use of an attentional model, as opposed to an LSTM. This, however, we propose to be simply a component of a more powerful method for detecting malware in cloud and networked structures.

\subsection{Level II: The Cloud: Networks and Malware Paths}
Now that we have latent representations of system logs, we can apply them in a wider context of a network. In this discussion, we have a single cloud with multiple interlinked VMs and other components, all of which we refer to as elements. Let $V$ be the set of elements in the cloud, and let $E$ be the set of edges, where an edge exists between two elements in the cloud if and only if there exists some relationship that, directly or indirectly, has a possible effect on malware spread. Each edge may have information associated with it (e.g., conditions that must be met for malware to spread from one vertex to another, temporal information, bandwith, latency) which can be represented as a real vector. Each vertex likewise may have information associated with it, as discussed in the previous subsection (i.e., system log information, metadata). 

We would like to perform several types of tasks at this stage.
\begin{enumerate}
    \item We would like to estimate the probability of a vertex being infected with malware given the topology of the network and the feature information of the vertices and edges.
    \item We would like to predict the spread of malware throughout the cloud at given timesteps, given a state and sequence of actions.
    \item We would like to discover actions that maximize the effectiveness of malware containment while minimizing the negative effects of system downtime and malware impact on the service capabilities of the cloud.
\end{enumerate}
We will investigate each of these tasks in this subsection. In general, we propose the use of graph neural network (GNN) architectures for capturing topological signal that may indicate the presence and characterize the behavior of malware in the cloud. We postulate that some clouds may have subnets or groups of systems with some shared characteristic (a hybrid cloud may have groups with different access privileges, some subset of the systems my have some library, etc.), and we pose this new construction as a hypergraph problem to which hypergraph learning may be applied. We also identify an open problem of discovering optimal actions that optimize certain criteria, and give an initial reinforcement learning-based approach.

\subsubsection{Inferring Malware Infection}
We are given a cloud that can be represented as a graph $G = (V,E)$ where each system $v$ in $V$ has an associated system log/metadata feature representation $r_v$ associated with it. A key intuition we leverage is that malware spread often displays characteristics of probabilistic epidemic/contagion models such as SIR \cite{sir}. Some malware programs may not attack immediately and instead lie inactive but present on infected systems, leading to the compartment-based  susceptible (S), delitescent (D) (not yet infective), infected (I), recovered (R) model proposed in \cite{websir}. Dynamic contact models such as these are useful heuristics, but each cloud is different and we would like to find a solution that captures the signal of a specific cloud to more effectively detect malware.

Here we draw on ideas from graph neural networks \cite{gnn, gcn, graphsage, GAT} and anomaly detection mechanisms on graphs \cite{netwalk, ding2019deep}. Letting $R$ be the matrix of features for vertices $r_v$, $A$ be the adjacency matrix of $G$, $D$ be the diagonal matrix where $D_{i,i} = \sum_{j} (A+I)_{i,j}$, and $f(R, A)$ be the function we would like to find that we would like to learn, the $l+1^{\text{st}}$ graph convolutional layer is defined as $$H^{(l+1)}= \sigma\!\left(D^{-\frac{1}{2}} (A+I)D^{-\frac{1}{2}}H^{(l)} W^{(l)} \right)$$ where $H^{(0)}=R$. This model captures the $(l+1)$-hop neighborhood topology of a vertex when thinking about it in a message-passing sense. Variations, such as sampling and aggregation found in GraphSAGE or multi-headed attention found in Graph Attention Networks, can also be used. This model is used to create effective embeddings that represent the vertices with respect to the network's topology and features, which make it ideal for understanding malware infections. 

If training data is available (for instance, one may be able to pretrain on and gather priors from the DARPA IDS evaluation dataset \cite{darpa}), then one of these GNN models could be trained to evaluate the state of each vertex in the network. The task could be, for instance, an indicator of whether or not the system has been infected ($\bold{y}\in \{0, 1\}$), a different categorical task identifying the type of malware present, or a regression task evaluating the Risk, Exploitability, and/or Impact scores of each system.

If training data is not available, malware detection is still tractable as an anomaly detection problem. Previously, the loss function is defined with respect to some task using squared-error (or similar) loss on the labels. Anomaly detection in attributed networks can performed without having access to labels, however, as was demonstrated in \cite{ding2019deep}. The loss function is instead based on a reconstruction of the structure of the network or attributes. In particular, the task is to rank all the nodes according to the degree of abnormality, such that the nodes that differ singularly from the majority reference nodes should be ranked highly. We now adopt a new reconstructive loss function, with $\hat{A}$ the estimated adjacency matrix, $\hat{R}$ is the estimated attribute matrix, and $\alpha$ is a hyperparameter: $$\mathcal{L}:=(1 - \alpha)||A - \hat{A}||^2_F + \alpha||R - \hat{R}||^2_F.$$ This directly gives us a means of determining the anomaly score of a vertex, given its neighborhood topology and its features: 
$$\texttt{anomaly}(v_i) = (1 - \alpha)||a_i - \hat{a}_i||_2 + \alpha||x_i - \hat{x}_i||_2$$
This procedure is, intuitively, used to find vertices that are out of the standard distribution of normally-operating (non-infected) systems. From this we can create a heatmap to discover areas of high vulnerability. 
\begin{figure}[h]
    \centering
    \includegraphics[width=230px]{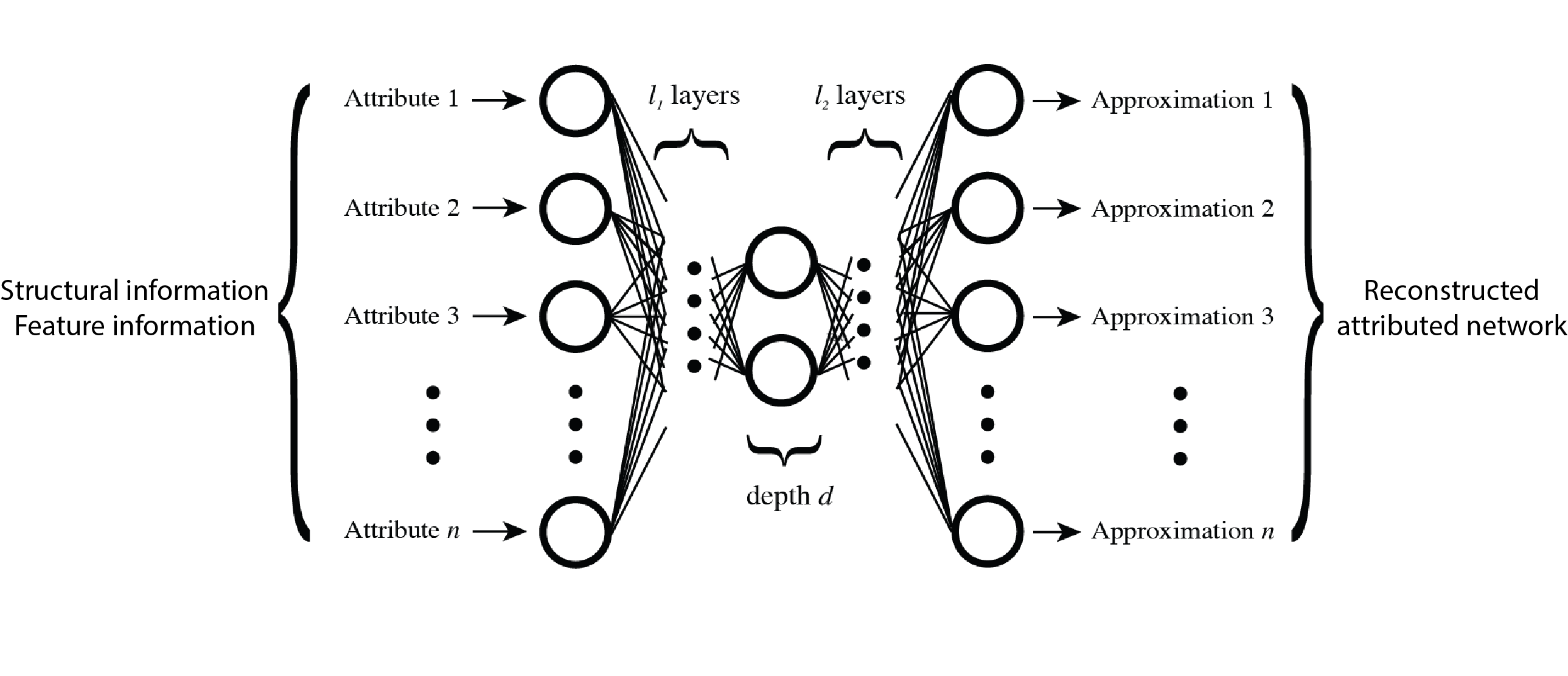}
    \caption{A deep graph autoencoder model is used in conjunction with reconstruction loss to find systems with anomalies in the cloud.}
    \label{fig:graphlstm}
\end{figure}
In certain cases, it may be more useful to perform inference on sets of systems rather than on systems or links between them. For instance, say that we'd like to ascertain the risk score of a particular subnet, in which many shared systems exist, the impact score of a specific library, which is present on several systems (along with other libraries), or the privacy level of a sector of a hybrid cloud, which shares devices with other sectors of differing privacy levels. In this case, we have not a graph problem but a hypergraph problem on our hands---permutation-invariant sets form hyperedges (the subnet, the library, etc.) while vertices still map to systems. For brevity, we note that the methods proposed for hypergraphs \cite{dhne, dhe, hgnn, sbai} are similar to those proposed for graphs, and the above discussion applies here as well. In this case we can perform classification or regression tasks to find, for instance, the probability that a subnet has been infected by malware, or the exploitability score of a specific library given the hypergraph topology and features of the systems within the hyperedge (using the library).

A final note with these methods is that they are inductive by nature; that is, we can learn on one graph and be able to generalize this knowledge to another unseen graph effectively. This is particularly important here, as the topology of the network would likely change over time as connections are created and broken between systems.

\subsubsection{Predicting Malware Spread}
We've discussed several methods for identifying vertices that have a high probability of being infected with malware in a network. How can we use this approach to predict how malware will spread? This is an important step in the containment of malware in the cloud. We propose the use of a recurrent or attention-based model for temporal networks. More specifically, given the vertex embeddings at time $1, 2, \ldots, t$, we would like to determine a vertex embedding at time $t+1, t+2, \ldots, t+k$, effectively predicting how the malware will spread. 

Direct methods for temporal embedding have been defined, which we do not go into detail for here. Another approach would be to embed the graph using the above methods, and apply a sequence completion task on these embeddings via an LSTM, Transformer, or other sequence model.

\begin{figure}[h]
    \centering
    \includegraphics[width=230px]{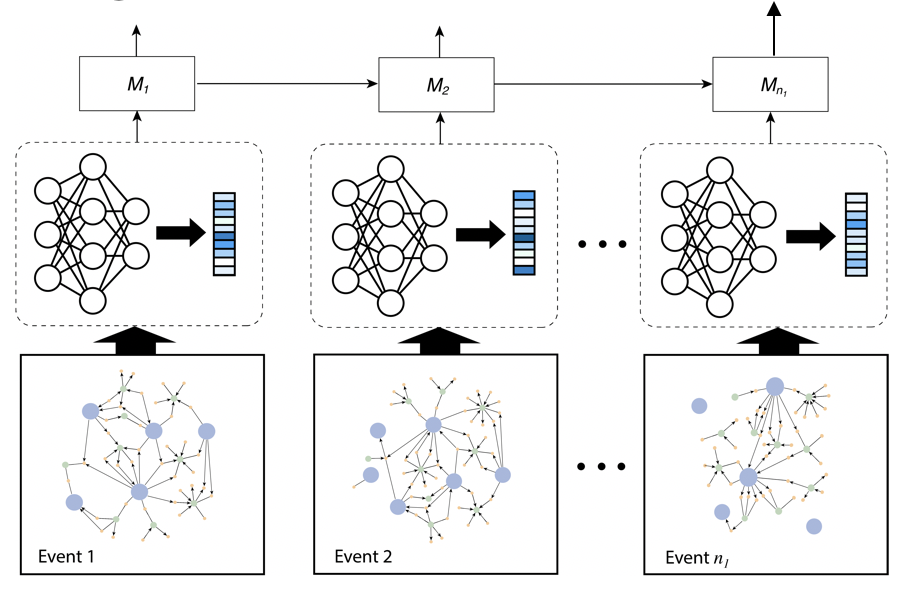}
    \caption{An LSTM is used as a sequence model on graph embeddings to predict future embeddings which indicate the spread of malware in the cloud.}
    \label{fig:graphlstm}
\end{figure}

Recurrent and temporal models for graph behavior are still an active area of research, so any new advances will likely improve the capabilities of this system to predict the spread of malware. The task specifically will be to ``color" the graph based on the Risk, Exploitability, or Impact scores as a result of malware spread at future timesteps. 

\subsubsection{Strategic Actions for Containment}
Once we've been able to evaluate the current state of the cloud with respect to malware and have predicted possible spreading patterns with the above models, our next task is to optimally contain it. Particularly, we need to minimize the cost incurred by taking an action (such as removal) on a vertex while minimizing impact of malware and observing the hard constraints imposed by the cloud. 
This definition is extremely broad; intentionally so, as different clouds have many different objectives, parameters, and priorities. The action space and method for selection is a difficult problem that we leave as future work largely. One initial approach would be to formulate it as a reinforcement learning problem, where the reward function is dense and defined by the above optimization problem. If the network is small enough and the actions are well-defined, a combinatorial optimization approach using the above objective function would suffice.






\subsection{Level III: Multicloud}
Many organizations and services rely on multiple clouds for various subservices and components. However, most commonly, the clouds and systems within them are not accessible to every participant in the multicloud. Organizations often prefer to keep their information, systems, and broader cloud private while still participating in the multicloud. How can we detect and contain malware effectively within the multicloud in a way that effectively draws knowledge and signal from the participating clouds without divulging specific non-disclosable information from these clouds?

We propose the use of federated learning in order to collect signal characterizing the presence and spread of malware from multiple individual clouds without creating the need to reveal specific network or system log information. 

\begin{figure}[h]
    \centering
    \includegraphics[width=250px]{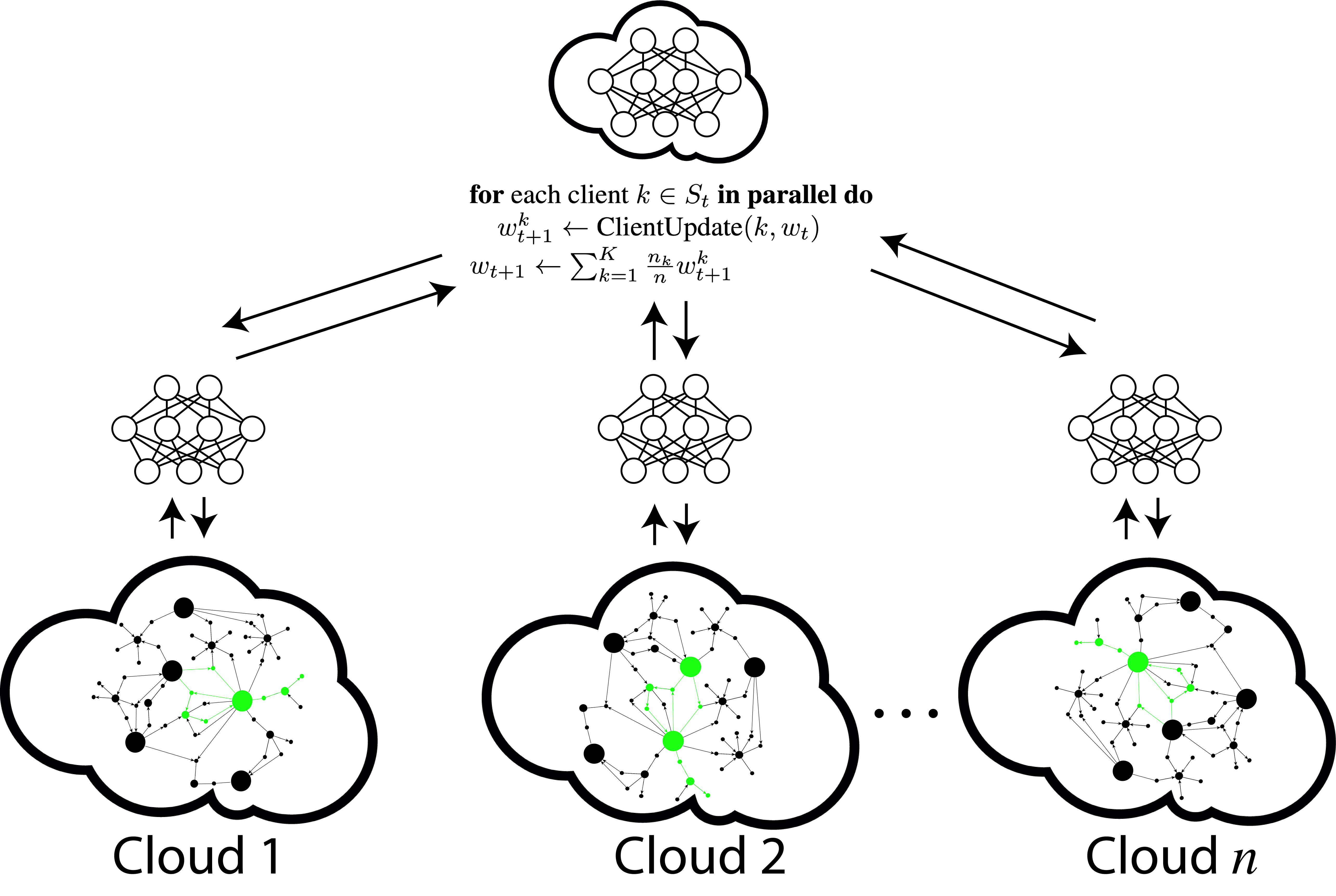}
    \caption{Federated learning for privacy-preserving malware detection and containment. Green vertices represent the path propagation of malware. Formula shown is a step in the Federated Averaging algorithm \cite{mcmahan2016communicationefficient}.}
    \label{fig:fedcloud}
\end{figure}

Suppose we have an end-to-end model $M$ that we want to use to perform some inference related to malware in a cloud (this inference task can be, for instance, malware detection, malware spread prediction, or containment/eradication decision-making). The goal of federated learning, in this case, is to learn the parameters of $M$ from system data stored across some number of clouds. In round $t \geq 0$, the server distributes the current model $M_t$ to a subset $S_t$ of $n_t$ clients. These clouds independently update $M_t$ based on their local data. Let the updated local models be $M^1_t, M^2_t, \ldots, M^{n_t}_t$. The update of cloud $i$ can be written as $H^i_t := M^i_t - M_t$ for $i \in S_t$.

The authors of \cite{fl2} note that these updates could be a single gradient computed from the cloud, but typically will be the result of a more complex calculation, such as multiple steps of stochastic gradient descent taken on the cloud's network. In any case, each selected cloud then sends the update back to the central aggregation service, where the global update is computed by aggregating all the clouds' updates using an aggregation algorithm such as the Federated Averaging algorithm \cite{mcmahan2016communicationefficient}. This procedure is illustrated in Figure \ref{fig:fedcloud}.

The privacy implications of this procedure have been studied formally as random noise $Z$ is added to the updates \cite{Abadi_2016, geyer2017differentially, fl3}. Much of this analysis relies on work done in differential privacy \cite{dp}. Let $M$ be a randomized mechanism mapping from a domain $D$ to a range $R$. $M$ satisfies $(\epsilon, \delta)$-differential
privacy if for any two adjacent inputs $d$, $d' \in D$ and for any subset of outputs $S \subseteq R$:
$$\mathbb{P}[M(d) \in S] \leq e^{\epsilon} \mathbb{P} r[M(d') \in S] + \delta.$$

To hide a single client’s contribution within the aggregation, the authors of \cite{geyer2017differentially} introduced  new central model $M_{t+1}$ is allocated by adding this approximation to the current central model $M_{t}$. 
\[
  M_{t+1} = M_t + \frac{1}{m_t}\bigl(
    \sum_{k=0}^{n_t} \triangle M^k / \text{max}(1,\frac{\Vert \triangle M^k \Vert _2}{S}) + \mathcal{N}(0,\sigma^2 S^2)
     \bigl)
\]

To achieve $(\epsilon, \delta)$-differential privacy in federated learning, the authors made use of the moments accountant as proposed in \cite{Abadi_2016}. Each time the central aggregator allocates a new model, an accountant evaluates $\delta$ given $\epsilon$, $\sigma$, and $m_t$. Training is then stopped once $\delta$ reaches a certain threshold, the choice of this threshold depending on the total number of clouds.

This is a particularly important guarantee for clouds that are private but vulnerable to malware. Cooperation and buy-in from more clouds in a multicloud environment is highly important in well-understanding the spread of malware in the ecosystem.
\section{Open Problems}\label{open}
We've discussed several approaches in detecting and containing malware in cloud environments. However, a number of open and difficult problems remain, which we identify and discuss here. It's important to note that the proposed approaches to solving these open problems are only at their most nascent stage; these are very much  ``blue sky" problems. However, they're important problems that if solved could have a significant impact on malware detection and cloud computing in general.
\begin{enumerate}
    \item Autonomous architecture of malware-resistant clouds: we've been able to analyze the spread of malware within a cloud. Can we then use this knowledge to make design decisions in cloud infrastructure, connectivity, and behavior which optimize certain properties while remaining robust against malware? An initial approach would be to again formulate this as a graph learning problem. One could use the GraphRNN \cite{graphRNN} model to generate a graph that meets these criteria and optimizes the desired properties. This is a difficult and wide-open problem.
    \item Objective functions for optimal decision-making: we defined a very broad objective function for penalizing actions taken on systems while also penalizing the presence of malware in these systems, subject to constraints imposed by the cloud requirements. This was left intentionally general, as requirements and objectives may vary widely among clouds, with the commonalities being a penalty on actions taken on devices and the presence of malware. We leave the in-depth definitions of this objective function for specific clouds as an open problem.
    \item Action space definition and subsequent action selection: given the state of a cloud, how can we act optimally so as to contain or eradicate malware inside? We proposed a reinforcement learning problem with a reward function defined by the objective function aforementioned, but this is somewhat of a ``placeholder" solution, as a supervised approach would be more tractable and well-defined. Identifying actions and the mechanisms for deciding which action to take given a cloud's state is left as an open problem.
    \item Open protocols and data for learning priors: cloud data is often sensitive and private, and for organizations without access to large clouds or multicloud environments, any sort of supervised learning approaches would be difficult to implement. One example for prior-learning we mentioned was system log data from the DARPA IDS evaluation dataset \cite{darpa}. Do there exist examples from the internet graph or other publicly available data that can assist in pretraining the models proposed? This is left as an open problem.
\end{enumerate}
\section{Conclusion}\label{conclusion}
In this paper, we introduced a hierarchical approach to performing malware detection and management using several recent advances in deep learning. We analyzed individual systems, inspecting and understanding their behavior by learning natural language understanding functions on their system logs via attention-based models such as the Transformer and BERT. Given a feature representation of each systems' logs using this procedure, we constructed an attributed network of the cloud with systems and other components as vertices and proposed an analysis of malware presence and spread through this network with inductive graph and hypergraph neural network models such as Graph Convolutional Networks, GraphSAGE, Graph Attention Networks, and Deep Hyperedges. We also proposed a general optimization problem which is used to discover optimal actions to take in handling malware within the cloud. With this foundation laid, we considered the multicloud case, in which multiple clouds with privacy requirements cooperate against the spread of malware, proposing the use of federated learning to perform inference and training while preserving the clouds' individual privacy. Finally, we discussed several open problems that remain in defending cloud computing enviroments against malware.
\newpage

\bibliographystyle{IEEEtran}
\bibliography{IEEEabrv,main.bib}

\end{document}